\documentclass[11pt,a4paper]{article}
\usepackage[hyperref]{emnlp-ijcnlp-2019}
\usepackage{times}
\usepackage{CJKutf8}
\usepackage{float}
\usepackage{latexsym}
\usepackage{graphicx}
\usepackage{amsmath}
\usepackage{multirow}
\usepackage{multicol}
\usepackage{color}
\usepackage{algorithmic}
\usepackage{algorithm}
\usepackage[font=small]{caption}
\graphicspath{ {./images/} }

\usepackage{url}

\aclfinalcopy 

\author{Kuan Fang \\
  Zhihu Search  \\
  {\tt fangkuan@zhihu.com} \\\And
          Long Zhao \\
  Zhihu Search \\
  {\tt zl@zhihu.com}  \\ \And
  Zhan Shen \\
  Zhihu Search \\
    {\tt shenzhan@zhihu.com} \\ \AND
  RuiXin Wang \\
  Zhihu Search \\
  {\tt wangruixin@zhihu.com} \\ \And
      RiKang Zhou \\
  Zhihu Search \\
  {\tt zhourikang@zhihu.com} \\ \And
        Liwen Fan\thanks{The work was done when the author was with Zhihu Search} \\
   {\tt levyfan@163.com} \\ 
  }
 
 \title{Beyond Lexical: A Semantic Retrieval Framework for Textual Search Engine}

\begin{document}
\maketitle
\begin{CJK*}{UTF8}{gbsn}
\begin{abstract}
  Search engine has become a fundamental component in various web and mobile applications. Retrieving relevant documents from the massive datasets is challenging for a search engine system, especially when faced with verbose or tail queries. In this paper, we explore a vector space search framework for document retrieval. Specifically, we trained a deep semantic matching model so that each query and document can be encoded as a low dimensional embedding. Our model was trained based on BERT architecture. We deployed a fast k-nearest-neighbor index service for online serving. Both offline and online metrics demonstrate that our method improved retrieval performance and search quality considerably, particularly for tail queries. 
\end{abstract}

\section{Introduction}
\label{section introduction}
Search engine has been widely applied in plenty of areas on the internet, which receives a query provided by users and returns a list of relevant documents within sub-seconds, helping users obtain their desired information instantaneously. Numerous technologies have been developed and utilized in real-world search engine systems\cite{yin2016ranking}. However, the existing semantic gap between search queries and documents, makes it challenging to retrieve the most relevant documents from tens of millions of documents. Therefore, there is still a large proportion of search requests that can not be satisfied perfectly, especially for long tail queries. 
\newline
\indent A search engine system is usually composed of three main modules,
\begin{enumerate}
 \item[-] query understanding module
 \item[-] retrieval module
 \item[-] ranking module 
\end{enumerate} 
\indent The query understanding module first parses the original query string into a structured query object\cite{riezler2010query}. More specifically, the query understanding module includes several sub-tasks, such as word segmentation, query correction, term importance analyze, query expansion, and query rewrite, etc. After the query string was parsed, an index module accepts the parsed query, and then retrieve the candidate documents. \newline 
\indent  We call this stage the retrieval stage or the first round stage. Most web-scale search engine systems use the term inverted index for document retrieval, where $term$ is the most basic unit in the whole retrieval procedure. In the first round stage, the retrieved documents are ranked by a simple relevance model, eg TF-IDF, BM25, and the top-N documents with the highest score are submitted to the next stage for ranking. Finally, the documents scored largest by a ranking function are returned to users eventually. \newline
\indent For a search system described above, the final retrieval performance is highly enslaved by these query understanding module. Take word segmentation as an example: this task segments raw continuous query string into a list of segmented terms. Since the word segmentation algorithm has the risk of wrong segmentation. If the error segmented term does not appear in the document space, then no document could be retrieved in the first round stage, and it will return a result page without any document which damages the user's experience seriously.  \newline
\indent There is a lot of work focused on better understanding queries to retrieve more relevant documents. However, since the final performance is influenced by all parts of the query understanding module. Attempts to optimize only one part is usually hard to contribute to a significant enhancement. To avoid the problems mentioned above, we propose a novel complementary retrieval system that retrieves documents without the traditional term-based retrieval framework. That is, instead of parse raw query into a structured query, we directly map both queries and documents into a low dimension of embedding. Then in the online serving, the k-nearest-neighbor documents of the given query in the latent embedding space are searched for retrieval. \newline 
\indent Recently, we have witnessed tremendous successful applications of deep learning techniques in information retrieval circle, like query document relevance matching \cite{Huang2013Learning} \cite{shen2014learning} \cite{shen2014latent}, query rewriting \cite{He2016Learning}, and search result ranking\cite {Haldar2018Applying}\cite{grbovic2018real}. However, it is still hard to directly retrieve relevant documents using an end2end fashion based on k-nearest-neighbor search in latent space, especially for long tail queries.   \newline
\indent The latest far-reaching advancement in natural language processing with deep learning, BERT\cite{devlin2018bert}, provides a turning point to make end2end retrieval realizable. In this paper, we present a document retrieval framework as a supplement to the traditional inverted index based retrieval system. We design a new architecture to retrieve documents without a traditional term-based query understanding pipeline, which avoids performance decay by each subtask of query understanding. We use BERT architecture as the general encoder of query and document strings, then we fine-tuned the pre-trained BERT model with human annotated data and negative sampling technique. Finally, we conduct both offline and online experiments to verify our proposed method.
To sum up, our main contributions are described below:
\begin{enumerate}
    \item We design a novel end2end document retrieval framework ，which is a supplement to traditional term-based methods.
    \item Our model is trained on transformer architecture, and a series of training techniques are developed for performance enhancement. 
    \item The proposed techniques can not only be used in document retrieval but also have a significant improvement for search ranking.
\end{enumerate}

The rest of the paper is organized as follows. We concisely review the related work in Section \ref{section related work}. Sections \ref{section approach} mainly describes our proposed methods. Offline and online experiments are detailed given in Section \ref{experiments} and Section \ref{online evaluate} respectively. Finally, we conclude and discuss future work in Section \ref{conclusion}.

\section{Related Work}
\label{section related work}
\subsection{query understanding}
There is a variety of work on search query understanding\cite{Prakash2014Techniques}, including query correction\cite{chen2007improving}, query term weighting\cite{zheng2015learning}, query expansion\cite{azad2017query} and query reformulation\cite{buck2017ask}. In general, these kinds of methods coherently rewrite the raw query into a new query, by replacing, adding, or removing terms or phrases in the raw query. The rewritten query gets better expression and therefore can retrieve more relevant documents than the original one.

\subsection{knn approximate \& text embedding}
Besides the inverted index, vector search engines\cite{gionis1999similarity} have also been widely applied in many information seeking tasks, like image search\cite{Ji2014Deep} and recommendation system\cite{45530}. \newline 
\indent To retrieve documents using a vector search, we need to map a piece of text into a low-dimensional numerical vector. Various embedding techniques have been developed and proven to have the powerful capability of capturing the semantic meaning of natural language text\cite{mikolov2013distributed}, \cite{pennington2014glove} \cite{kusner2015word}. However, these kinds of models are still not capable of complicating text encoding, especially for long tail text queries.

\subsection{deep matching}
More recently, researchers have been describing the various architecture of neural models\cite{mitra2017neural}. In text relevance matching area, we can divide most models into two typical categories, namely representation\cite{Huang2013Learning} based models and interaction based models\cite{pang2016text} \cite{xiong2017end} \cite{dai2018convolutional}. The representation models , like DSSM, are trained to obtain high-level representations of query and document respectively, then use vector distance between the query and document embedding for text relevance score. While the interaction based models first compute the term correlation matrix between query and documents and calculate semantic matching similarity based on the correlation matrix. Both representation models and interaction based models could be trained from massive click feedback data\cite{joachims2002optimizing}\cite{agichtein2006improving} or industrial annotation.  These two kinds of model architecture are broadly deployed in real-world search engine systems, especially in ranking phase. For the representational models, once we obtained the high-level representation of raw texts, we can retrieve documents through the k-nearest-neighbor space search. However, the performance of representation based models are usually poorer than interaction based models, which makes k-nearest-neighbor retrieval hard to deploy in the real-world systems, since too many irrelevant documents retrieved may even damage overall performance. \newline
\indent \cite{zamani2017relevance} developed an architecture to transform the text into a sparse representation, while they still retrieve documents using a term-based index like lucene\footnote{https://lucene.apache.org/} because the non-zero value in the sparse representations is treated as virtual terms. \cite{bai2018scalable} \cite{grbovic2016scalable} developed a uniform query and document embedding framework by generating ngram embedding using user session and click data, and then generalize it to arbitrary text by mean average pooling of ngram embedding. Since ngram is a common and effective skill in a variety of NLP tasks, training a good ngram representation requires a massive of datasets, which may be a bottleneck for many researchers and companies. Meanwhile, the model capacity of DSSM and its' variations makes it not capable to capture complex semantic meanings of natural language.\newline
\indent Recently, ELMo\cite{peters2018deep}, GPT-2\cite{radford2019language} and BERT\cite{devlin2018bert} show the great power of unsupervised pre-training in NLP tasks. The BERT model is built on a 12 layer transformer architecture, pre-trained with large scale text data. The pre-trained models can be fine-tuned easily and outperform many state-of-art models in various NLP tasks. We used the pre-trained BERT-Base(Chinese)\footnote{https://github.com/google-research/bert} model released by Google and fine-tuned the model for semantic representation. Our fine-tuned model outperformed many state-of-art models in deep relevance matching, and obtain a great success in semantic retrieval task.

\section{Approach}
\label{section approach}
In this section, we first illustrate our proposed semantic retrieval framework, which is composed of both offline and online parts respectively. Then, we introduce the model structure used for encoding queries and titles, and the techniques we used to boost the performance.
\begin{figure}[t]
\includegraphics[width=8cm]{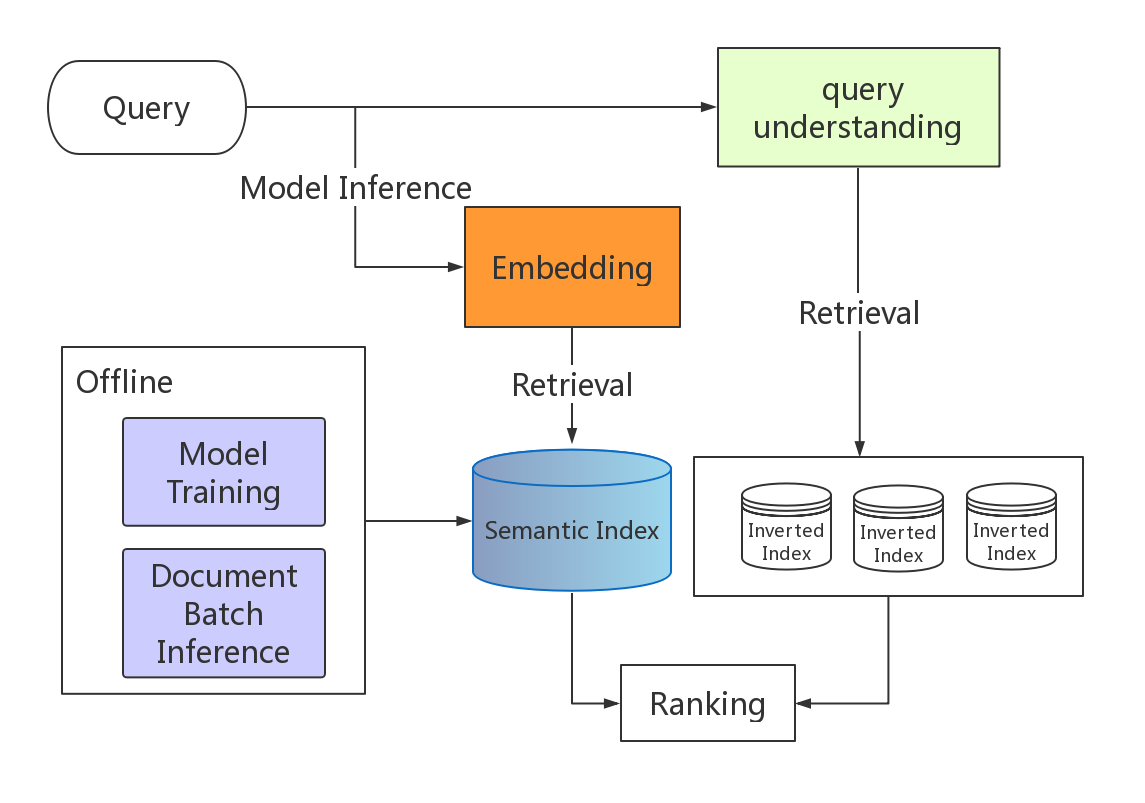}
\centering 
\setlength{\abovecaptionskip}{0pt}%
\caption{Overall framework of our proposed work.}
\label{fig 1}
\end{figure}
\subsection{Deep Semantic Retrieval Framework}
Figure \ref{fig 1} shows our proposed system architecture. The offline module includes model training, document embedding inference, and semantic index builder. While in online serving, both query's semantic embedding and traditional term base query parser are computed, and then those two results are sent to semantic index service and inverted index service respectively for document retrieval. Finally, documents retrieved from both index services are merged and sent to ranking service for document scoring. 
\subsection{Deep Semantic Representation Model}
The pre-trained BERT model can be leveraged for semantic ranking and matching\cite{qiao2019understanding} in various ways. We developed two models here: BERT(rep) and BERT(rel). The BERT(rep) model uses the pre-trained BERT model to obtain embedding of query and doc respectively, while the BERT(rel) model concatenates query and document first and get the one representation for a query document pair. The final score of a query, document pair is computed as below:
\begin{equation} \small
\label{bert-rep}
    BERT(rep)(q, d) = dot(\frac{1}{L}\sum_{i=1}^L\Vec{q}_i^{last},\frac{1}{L}\sum_{i=1}^L\Vec{d}_i^{last})
\end{equation}
In the equation \ref{bert-rep}, we use the mean average of last layer as encoder output for each query and document, and compute the dot product of two embedding as matching score, where $L$ represents the max sequence length. We also tried directly using the last layer of [CLS] term's embedding, but performed worse than the average pooling described in equation \ref{bert-rep}.
\begin{equation} \small
\label{bert-rel}
    BERT(rel)(q, d) = \Vec{w}\times\Vec{qd}_{cls}^{last}
\end{equation}
The equation \ref{bert-rel} use embedding of last layer's [CLS] token and weighted sum it to a scalar by vector $\Vec{w}$, where $\Vec{w}$ is a full connection layer with only one output. The model capacity of this method is more powerful than BERT(rel) because it calculates the term interaction between the query and document in the self-attention layers. However, since the BERT(rel) model is an interaction base model, this model can not be applied to semantic retrieval. \newline 
\indent Both two models are trained through a supervised learning fashion, with a pairwise max margin hinge loss to distinguish relatively positive and negative samples. The loss function for one query is:
\begin{equation}
\small \frac{1}{M}\sum_i\sum_{j\neq i}max(0, \tau -(y_i-y_j)\times(p_i-p_j))
\end{equation}
where $p_i$ and $p_j$ represent to model score computed for each $<query, document>$ pair, and $y_i$ and $y_j$ is the label for each document respectively. $\tau$ is the hyper parameter called margin to determine how far the model need to push a pair away from each other.
The margin parameter is tuned for the best performance here.
\begin{table*}
\small
\centering
\caption{Examples of the dataset.  There are four titles correspond to query “机器学习 编程”(machine learning programming). They are annotated positive, annotated negative, $NEG_{global}$ and $NEG_{cluster}$ respectively.}
\label{table 1}
\setlength{\abovecaptionskip}{10pt}%
\setlength{\belowcaptionskip}{0pt}%
\begin{tabular}{l|l}
\hline
\hline
\textbf{Query} & Titles  \\ \hline
\\   &机器学习，在理论和编程方面要如何准备？ \newline
\\   & How to prepare programming parts to study machine learning 
\\ \cline{2-0}
\\ 机器学习 编程   &深度学习如何入门？\newline 
\\ Machine Learning Programming   & How to get started to study deep learning  
\\ \cline{2-0}
\\ & 什么是好朋友，有谁给过你怎样的深感动？ 
\\ &                      What is good friend, who has touched you deeply
\\ \cline{2-0}
\\ & 回忆我的编程之路  \newline
\\ & recall my history of programming
\\ \hline
\hline
\end{tabular}
\end{table*}
\subsubsection{Additive Sampling}
\label{sample desc}
We use the additive data sampling technique to further enhance model performance. Therefore, the data we used to train our model is comprised of two parts, human annotated data, and negative sampled data. Negative sampling has been successfully applied in many tasks, such as neural language modelling\cite{mikolov2013distributed}, e-commerce list embedding\cite{grbovic2018real}, graph embedding\cite{ying2018graph} and so on. \newline 
\indent Sampling negative training instance is also useful for model training in this scenario, since different from traditional term-based retrieval method, the vector space search is much more likely to retrieve irrelevant documents. Thus we propose to augment more irrelevant documents. When the negative samples were added to training, the model learned to push relevant and irrelevant documents away from each other, then the model is more robust to noisy documents. \newline
\indent A straightforward way of negative sample mining is to select negative samples corresponding to a uniform distribution over the whole corpus, in particular, irrelevant documents here. However, this simple strategy fails to generate hard negative samples, which provide more important information for the model. Therefore, we propose another negative sampling method. At first, we train a baseline model with only human annotated data. Then we use this model to encode documents and queries. After that, we use an unsupervised cluster algorithm to assign each document and query a cluster id. Finally, we uniformly random selected negative documents from the cluster that query was distributed. \newline 
\indent For convenient, we call this kind of negative sampling name of $NEG_{cluster}$, and globally sampled data name of $NEG_{global}$. We append $NEG_{cluster}$ and $NEG_{global}$ to the raw dataset for per query and fine-tuned the model again to obtain our final model.re   \newline
\begin{algorithm}[h]\small
\caption{\small Training Framework of our proposed model}
\label{algo 1}
\begin{algorithmic}[1]
\REQUIRE ~~\\ 
 human annotated data $D$,
 BERT pre-trained model $M$
\STATE $M_1 \gets \left\{D, M \right\}$, fine-tune the model $M$ by $D$
\STATE compute embedding $E$ for query and doc using $M_1$
\STATE compute cluster centroids $C$ by $E$
\FORALL {$d \in Docs $}
\STATE compute closest centroid $C_d$ for $d$
\ENDFOR
\FORALL {$q \in Query $}
\STATE compute closest centroid $C_q$ for $q$
\STATE uniform sample $NEG_{global}$ from whole doc set
\STATE uniform sample $NEG_{cluster}$ among docs where $C_d = C_q$
\STATE $D_{1}(q) = \left\{D(q) \cup NEG_{global} \cup NEG_{cluster}\right\}$
\ENDFOR 
\STATE $M_2 \gets \left\{D_1, M \right\}$, fine-tune the model $M$ by $D_1$
\ENSURE ~~ \\
 BERT model $M_2$
\end{algorithmic}
\end{algorithm}
\indent We show the whole training procedure in the Algorithm \ref{algo 1}, and 
Table \ref{table 1} illustrate examples of human annotated samples and auto-generated negative samples. Four titles are corresponding to each query, which represents the human annotated positive title, human annotated negative title, $NEG_{global}$ title and $NEG_{cluster}$ title respectively. From the table, we can see that the $NEG_{cluster}$ sample's meaning is much closer to query than that of $NEG_{global}$, which makes the model more robust for hard samples.
\subsection{Online Serving}
Once the model was trained, we need to serve it on the fly. We first computed the embedding of all documents and build a vector index using faiss\footnote{https://github.com/facebookresearch/faiss} \cite{JDH17}, which was open sourced by facebook and support k-nearest-neighbor search for vector data in milliseconds. We developed a c++ based semantic index server to provide efficient concurrent online service. Our model was inferenced on a GPU server, and inference speed was accelerated 2 times faster than tf-serving through a c++ based library developed by us. During the online serving, when a query was received, the GPU server first inferences the query embedding, and downstream sends the query embedding to semantic index service for document retrieval. For the balance of efficiency and effect, we retrieve k most similar documents in the semantic service for next stage ranking, where k is set to 20 here.

\section{Offline Experiments}
\label{experiments}
In this section, we carry out offline experiments to illustrate the performance of our proposed semantic retrieval methods. In the experiment, we train the model with 1 epoch, use Adam\cite{kingma2014adam} with a learning rate of $10^{-5}$, β1 = 0.9, β2 = 0.999.  
\subsection{DataSets}
\begin{table*}
\centering
\small
\caption{Brief statistics of annotated data}
\label{dataset}
\setlength{\abovecaptionskip}{0pt}%
\setlength{\belowcaptionskip}{15pt}%
\begin{tabular}{|l|l|l|l|l|l|}
\hline
   & \textbf{Query}  & \textbf{QueryDoc} & \textbf{Excellent} & \textbf{Fair}  & \textbf{Bad}\\ \hline
\textbf{TrainSet}  & 36159  & 1181229 & 106181 & 357552 & 717464 \\ \hline
\textbf{TestSet}   &  2703  & 84244   & 9552 & 17801 & 56891 \\ 
\hline
\end{tabular}
\end{table*}
The data annotated by human editors is a list of triplets like $<$query, doc, relevance$>$. The relevance score has three grade {0, 1, 2}, which represents $bad$, $fair$ and $excellent$ respectively. The dataset contains 36159 queries and 1181229 query doc pairs. Beside the dataset for training, we additionally annotated a small dataset for test, the test dataset contains 2703 queries and 84244 query-doc pairs. The summarize of dataset is shown at Table \ref{dataset}. 
\subsection{Evaluation Metrics}
We evaluate our proposed model from ranking and retrieval aspects. 
We compared the ranking performance using Normalized Discounted Cumulative Gain(NDCG), and retrieval performance with Recall. The way how these metrics are calculated will be introduced in  Section \ref{exp recall} and Section \ref{exp rank} respectively. 
\subsection{Baselines}
\begin{itemize}
    \item ClickSim \newline
    A relevance matching model\cite{jiang2016learning} which use web-scale click data to generate term representations for query and document, and use cosine similarity to represent query document relevance.
    \item K-NRM \newline
    An interaction based matching model using kernel pooling\cite{xiong2017end}. 
    \item Match Pyramid \newline
    An interaction based matching model using convolutions on term matching matrix\cite{Pang2016A}.
    \item DSSM \newline
    A representation based model proposed by Microsoft Research\cite{Huang2013Learning}. The model proposed here using word vectors pre-trained on document title corpus. And three full connection layer with size of 300, 300, and 128 dimensions are used for text encoding.
\end{itemize}
\subsection{Recall Performance}
\label{exp recall}
\begin{table}
\small
\caption{Recall performance of different models. The \textbf{Lexical} represents the traditional term-based retrieval.} 
\label{recall}
\centering
\setlength{\abovecaptionskip}{0pt}%
\setlength{\belowcaptionskip}{5pt}%
\begin{tabular}{|l|l|l|}
\hline
  \textbf{Methods}  & \textbf{Recall Num} & \textbf{Recall Rate}  \\ \hline
Lexical  & 12963 & 54.9\%  \\ \hline
DSSM   &  5  & n.a \\ 
DSSM+Lexical   &  12963  & 54.9\%   \\ \hline
BERT(rep)   &  9794  &  \textbf{41.5\%}   \\ 
BERT(rep)+Lexical   &  16394   & \textbf{69.4\%} \\ 
\hline
\end{tabular}
\end{table}
We use metric Recall to evaluate the model's retrieval performance here. This metric measures how many relevant documents are retrieved by a given model. For a given query $q$, the Recall rate is calculated as, 
  \newline
    \begin{equation} \small
        Recall_q = \frac{Ret_q \cap Rel_q}{Rel_q}
    \end{equation}
where $Ret_q$ represents the retrieved documents for $q$, $Rel_q$ stands for all the relevant documents for query $q$, where relevant documents are defined as document relevance annotated larger than 0 here. \newline 
\indent To evaluate the recall performance offline, we first built semantic index both for our model and baseline model. We computed representation for document title of each model, then we used the representation embedding to build semantic index. Once queries' embedding of each model were computed, we retrieved the top k documents by k-nearest-neighbor search. Besides comparing the recall measure of different models only using semantic index, we compared the recall enhancement when the semantic index was added to the lexical inverted index. We used a commercial term-based inverted index engine developed by us and build a lexical index with it. Both lexical inverted index and semantic index were built to retrieve documents, with top 300 and top 20 respectively. Then we calculated the recall of the union set. \newline
\indent In the experiment, since document size of testset is small, we need a larger document corpus to make the recall measured more accurately. Therefore, both semantic index and lexical index were built with all human annotated data, including trainset and testset. And recall metric were calculated using only queries in the testset.  \newline 
\indent Table \ref{recall} shows the result of different models, BERT(rep) outperforms baseline model DSSM significantly in the recall measure. And after adding our model as a supplement to the lexical index, the recall rate is improved from 54.9\% to 69.4\%. While the baseline model, DSSM performs poorly on this task. 
\subsection{Ranking Performance}
\begin{table*}
\small
\centering
\caption{Ranking performance between different models.} \label{ndcg}
\setlength{\abovecaptionskip}{0pt}%
\setlength{\belowcaptionskip}{5pt}%
\begin{tabular}{llll}
\hline
\hline
\textbf{Method}   & \textbf{NDCG@1}  & \textbf{NDCG@3} & \textbf{NDCG@5} \\ \hline
Match Pyramid  & 0.7332  & 0.7312 & 0.7425 \\ 
K-NRM    & 0.712   & 0.7118 & 0.7251 \\ 
ClickSim & 0.619   & 0.6164 & 0.6315 \\ \hline
BERT(rep) & \textbf{0.7775 (6.04\%)}  & \textbf{0.7754 (6.04\%)} &
\textbf{0.7849 (5.71\%)} \\
BERT(rel) & {0.8009}  & {0.7962} & {0.8044} \\
\hline
\hline
\end{tabular}
\end{table*}
\begin{enumerate}
\item[-] {NDCG score} \newline
\label{exp rank}
    Since our proposed model could not only be applied in document retrieval but also applied in the ranking stage. We measured the model's ranking quality through Normalized Discounted Cumulative Gain(NDCG).  For a ranked document list, the NDCG for a query is calculated as, 
    \begin{equation} \small
        NDCG_n = \frac{DCG_n}{IDCG_n}
    \end{equation}
    where $IDCG_n$ represents the $DCG$ score when the list was perfectly ranked by relevance. We compute following variation of Discounted Cumulative Gain(DCG)\cite{jarvelin2002cumulated},
    \begin{equation} \small
    \label{DCG}
        \small DCG_n = \sum_{i=1}^N\frac{2^{label_i}}{\log_2(i+1)}
    \end{equation} 
    According to the equation \ref{DCG}, higher relevance label contribute to higher weight in the computation. We calculate $NDCG$ with different rank list size of $\left\{1,3,5\right\}$ respectively.
    Table \ref{ndcg} shows that our model is superior to the state of art deep relevance matching models, and BERT(rep) model is slightly worse than the BERT(rel) model since BERT(rel) model uses self-attention between the query and title tokens before aggregates final score. However, both the BERT(rep) model and BERT(rel) model outperform other baselines significantly.
    \item[-] {feature importance in ranking model} \newline
                 \begin{table}[H]
        \small
         \centering
         \caption{Feature importance in GBDT ranking model} 
                \setlength{\abovecaptionskip}{0pt}%
         \setlength{\belowcaptionskip}{0pt}%
         \begin{tabular}{|l|l|}
         \hline
\textbf{FeatureName}     & \textbf{ImpFraction} \\ \hline
BERT(rep)       & \textbf{34.11\%}    \\ \hline
ClickSim        & 10.65\%   \\ \hline
K-NRM            & 4.72\%    \\ \hline
Match Pyramid    & 2.82\%    \\ \hline
         \end{tabular}
         \label{feat_imp}
\end{table}
    We feed the doc product of query doc embedding into a gbdt ranking model\cite{burges2010ranknet} as a relevance feature, and observe the feature importance after the tree model was trained. The feature importance was computed by the statistics collected during the tree ensemble training procedure. Table \ref{feat_imp} shows that without adding BERT(rel) feature, the BERT(rep) feature ranks first in the ranking function, and accounts for 34\% of importance among all features in our ranking function. 
     \end{enumerate}

\subsection{Analysis of Negative Samples}
In Section \ref{sample desc}, we described two negative sampling generator method: the $NEG_{global}$ samples and  $NEG_{cluster}$ for training data enhancement. We tuned the negative samples size, and obtained the best performance with 10 $NEG_{global}$ and 10 $NEG_{cluster}$ respectively. After adding negative samples, the average negative sample size for a given query increased from 19.9 to 39.9. Table \ref{sampling perform} shows the model performance with different kinds of negative samples. Only adding $NEG_{global}$ can improve NDCG@3 at about 0.5\%, when adding $NEG_{cluster}$ , the NDCG@3 is further improved by 0.8\%. Therefore, the overall measurements are enhanced by 1.4\% after additive sampling.
        \begin{table}[H]
        \small
                          \caption{Model performance with different sampling, where HM represents the human annotated data. We show the NDCG@3 here because the NDCG of other rank performs similarly.} 
                           \label{sampling perform}
         \centering
                \setlength{\abovecaptionskip}{10pt}%
         \setlength{\belowcaptionskip}{5pt}%
         \begin{tabular}{|l|l|}
         \hline
\textbf{Model}     & \textbf{NDCG@3} \\ \hline
HM       & 0.7615  \\ \hline
HM + $NEG_{global}$        & 0.7669   \\ \hline
HM + $NEG_{global}$ + $NEG_{cluster}$   & \textbf{0.7754}   \\ \hline
         \end{tabular}

\end{table}
\subsection{Results of different pooling method of BERT}
\begin{figure}[t]
\includegraphics[width=7cm]{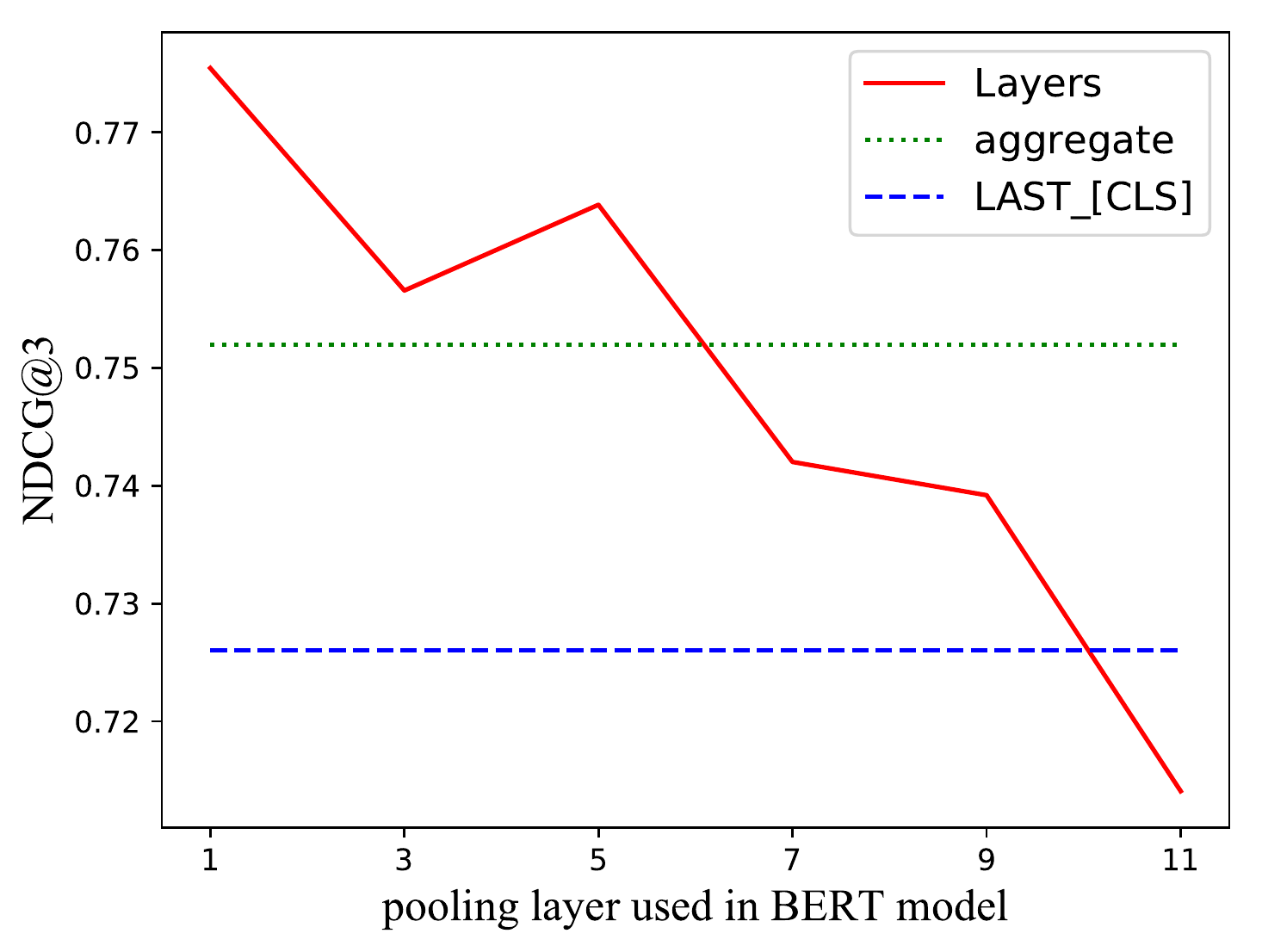}
\centering
\caption{Affect of using different layers and pooling method.  The x-axis represents the layer used for text pooling, starting from 1 to 11, and the y-axis represents the NDCG@3 metric.}
\label{pooling layer}
\end{figure}

In this paper, we use the reduce-mean of the last layer as BERT(rep) model's pooled output. Different layers of BERT may own different aspects of knowledge about the input sequence. To verify the effectiveness of different layers, we trained different models, with pooled output from different layers respectively. From Figure \ref{pooling layer}, the red solid line shows that the layer closest to last obtains higher NDCG measure. This is reasonable since higher layers make the model contains more parameters. \newline 
\indent Besides comparing the results of different layers, we also developed a method aggregate the embedding of all layers. In this method, an attention layer calculates the weight across different layers, therefore a weighted sum of each layer's embedding on each position is the final representation of each term. After that, we used reduce-mean of all terms' embedding as the final pooled output. The result of aggregation is shown as the green dot line, which does not outperform simple average pooling on the last layer. Meanwhile, we also tried using [CLS] term's embedding of the last layer as pooled output, but it behaved even worse. In conclusion, using mean average pooling of the last layer as final pooled output performs best in this scenario, even though some work claims aggregating layers is useful \cite{kondratyuk201975}.

\section{Online Evaluation and Case Study}
\label{online evaluate}
\subsection{Online A/B Testing}
After offline evaluations, we conduct an online a/b test to further verify our proposed system. In the online experiment procedure, 40 percent of online traffic were randomly distributed to four groups, 2 control groups, and 2 experimental groups. The metric we used to evaluate is the Clicked Search Rate(CSR), which is computed as: 
   \begin{equation} \small
        CSR = \frac{SearchNum_{clicked}}{SearchNum}
    \end{equation}
After a week's observation, as shown in Table \ref{online csr}, the overall CSR of two experimental groups both surpass two control groups by 0.65\%, which is relatively a huge improvement to our experience. We also examined the online performance for queries with different frequency. We split queries into Top, Torso, and Tail by query search times in a day. Since our proposed method mainly focuses on boosting the performance of long tail queries, we can see the CSR metric is not significant in the Top and Torso query part. But the metric increased by nearly 1.05 \% in the Tail part, which contributed to the largest algorithm iteration in the first half of 2019. 
        \begin{table}
                \small

                          \caption{Clicked Search Rate(CSR) of experimental groups and control groups. We set two control groups and two experimental groups to eliminate the online traffic bias.} 
                                  \setlength{\abovecaptionskip}{10pt}%
         \setlength{\belowcaptionskip}{0pt}%
        \label{online csr}
         \centering

         \begin{tabular}{|l|l|l|l|l|}
         \hline
\textbf{Group}     & \textbf{Total}  & \textbf{Top}  & \textbf{Torso}  & \textbf{Tail} \\ \hline
Control-1       &  76.74\% & 75.52\% & 80.32\%	 &74.97\%	  \\ 
Control-2       &  76.74\% & 75.493\% & 80.37\%	 &74.98\%	  \\ \hline
Exp-1        & \textbf{77.30\%} & 75.46\%&  80.35\%	 & \textbf{76.05\%}	 \\ 
Exp-2        & \textbf{77.31\%} & 75.53\%&  80.38\%	 & \textbf{76.03\%}	 \\ \hline
         \end{tabular}

\end{table}
\subsection{Case Study}
This section highlights some good cases after our system was deployed online. \newline 
\indent We show the final result ranked at top 6 for query “送外卖不认识路” (do not know the way to deliver food) at Table \ref{showcase}, where SEMANTIC represents the document retrieved from the proposed semantic index, and LEXICAL for traditional term-based inverted index. \newline
\indent In this case, three documents are retrieved from semantic index, and the relevance is also much better than the document from traditional inverted index. Notice that there are many ways to express “不认识路”(do not know the way) in Chinese, while the semantic index retrieved documents indeed capture the several alternatives of expressing it:  “不知道路线”, “不认路”, “不懂路”. And the term retrieved document only contains the same term "不认识路" as query expressed. 
\begin{table}[H]
\caption{Top ranked titles for query “送外卖\textcolor{blue}{不认识路}” (do not know the way to deliver food)}
\label{showcase}
\small
\centering
\setlength{\abovecaptionskip}{3pt}%
\setlength{\belowcaptionskip}{0pt}%
\begin{tabular}{|l|l|}
\hline
\textbf{Index}    & \textbf{Title}               \\ \hline
& \\
SEMANTIC & 配送外卖\textcolor{red}{\textbf{不知道路线}}怎么办？       \\ 
&  What if the food delivery  \\
& do not know the deliver route? \\ \hline
& \\
LEXICAL     & 本人是送外卖新手，不知道其中的 \\
& 送餐技巧 \\ 
& I am new to food delivery, \\ 
& and I do not know the deliver skills. \\ \hline
& \\
SEMANTIC & 为什么外卖员\textcolor{red}{\textbf{不认路}}？     \\ 
&  Why does the food delivery do not know  \\
& deliver route? \\ \hline
& \\
LEXICAL     & 恶劣天气叫外卖的行为是否恰当？     \\ 
& Is the behavior of takeaway in bad  \\ 
& weather appropriate? \\ \hline
& \\
LEXICAL & 我想送外卖，又怕\textcolor{blue}{不认识路}。\\ 
& I want to be a food delivery, \\
& but I am afraid I don't know the way. \\ \hline
& \\
SEMANTIC & \textcolor{red}{\textbf{不懂路}}可以送美团外卖吗 \\
& Can I be a food delivery at MeiTuan \\ 
& if I am not familiar with route? \\ \hline
\end{tabular}
\end{table}

\section{Conclusion}
\label{conclusion}
In this paper, we present an architecture for semantic document retrieval. In this architecture, we first train a deep representation model for query and document embedding, then we build our semantic index using a fast k-nearest-neighbor vector search engine. Both offline and online experiments have shown that retrieval performance is greatly enhanced by our method. \newline
\indent For the future work, we would like to explore a more general framework that could use more signals involved for semantic retrievals, like document quality features, recency features, and other text encoding models.

\bibliography{semantic}
\bibliographystyle{acl_natbib}
\end{CJK*}
\end{document}